\documentclass[reprint,floatfix,superscriptaddress,nofootinbib]{revtex4-1}
\usepackage{graphicx}
\usepackage[product-units = power]{siunitx}
\usepackage[version=3]{mhchem}
\usepackage[utf8]{inputenc}
 \usepackage{setspace}
 \usepackage{fixmath}
 \usepackage{ulem}

\DeclareSIUnit\BohrMagneton{$\mu_\textrm{B}$}
\DeclareSIUnit\formulaunit{f.u.}
\DeclareSIUnit\atomicunit{a.u.}
\DeclareSIUnit\arbunit{arb.unit}
\DeclareSIUnit\torr{Torr}

\begin{document}

\title{Spin-orbit torque switching without external field with a ferromagnetic exchange-biased coupling layer}
\author{Yong-Chang Lau$^\dagger$}
\author{Davide Betto$^\dagger$}
\email[Corresponding author: ]{bettod@tcd.ie}
\author{Karsten Rode}
\author{J. M. D. Coey}
\author{Plamen Stamenov}
\affiliation{CRANN, AMBER and School of Physics, Trinity College Dublin, Dublin
2, Ireland}

\def\@fnsymbol#1{\ensuremath{\ifcase#1\or *\or \dagger\or \ddagger\or
   \mathsection\or \mathparagraph\or \|\or **\or \dagger\dagger
   \or \ddagger\ddagger \else\@ctrerr\fi}}

\renewcommand*{\thefootnote}{\fnsymbol{footnote}}
\footnotetext[2]{ These
authors contributed equally to this work
}

\begin{abstract}

Magnetization reversal of a perpendicular ferromagnetic free layer by
spin-orbit torque (SOT)\cite{Miron2011,Liu.PRL.2012a,Liu.Science.2012,Fan.NatMat.2014} is an attractive alternative to
spin-transfer torque (STT) switching\cite{Slonczewski1996,Berger.PRB.1996} in
magnetic random-access memory (MRAM) where the write process involves passing
a high current across an ultrathin tunnel barrier\cite{Ikeda2010}. A small
symmetry-breaking bias field is usually needed for deterministic SOT switching but
it is impractical to generate the field externally for spintronic applications. Here, we demonstrate robust
zero-field SOT
switching of a perpendicular \ce{Co90Fe10} (\ce{CoFe}) free layer where the symmetry is broken by
magnetic coupling to a second in-plane exchange-biased \ce{CoFe}
layer via a nonmagnetic \ce{Ru}
spacer\cite{Parkin1990}. The preferred magnetic state of the free layer is determined by the current polarity and the nature of the
interlayer exchange coupling (IEC). Our strategy offers a scalable solution to
realize bias-field-free SOT switching that can lead to a generation of
SOT-based devices, that combine high storage density and endurance with potentially low power consumption.

\end{abstract}
\maketitle

For memory applications, a storage layer with perpendicular magnetic anisotropy
(PMA) is preferred because it offers higher
storage density, better thermal stability and lower power consumption
than a layer with easy plane anisotropy\cite{Mangin2006,Ikeda2010}. However,
deterministic SOT switching of a perpendicularly magnetized nanomagnet usually relies on an
external magnetic field to break the symmetry\cite{Miron2011,Liu.PRL.2012a,Liu.Science.2012}. SOT switching without an external field has recently
been demonstrated in systems with lateral asymmetry\cite{Yu2014} or with tilted magnetic
anisotropy\cite{You.Arxiv.2014} but neither of these schemes is easily
scalable\cite{note_on_Fukami}.

Here, we combine two concepts that have been developed in the context of modern hard-disc read heads and magnetic tunnel
junctions: exchange bias\cite{Nogues1999} and exchange coupling across a thin spacer\cite{Parkin1990} to
achieve scalable 
SOT switching without an external field.
Using a stack based on a perpendicularly-magnetized \ce{Co90Fe10} (\ce{CoFe})
free layer sandwiched between a \ce{Pt}
underlayer and a \ce{Ru}
overlayer, we show that the free layer can be deterministically switched by SOT from the \ce{Pt}. The symmetry-breaking issue is
resolved
by exchange coupling the free layer, via a \ce{Ru} spacer, to an in-plane 
exchange biased \ce{CoFe} pinned layer.

\begin{figure*}[ht]
\begin{center}
\includegraphics[
width=1.0\textwidth]{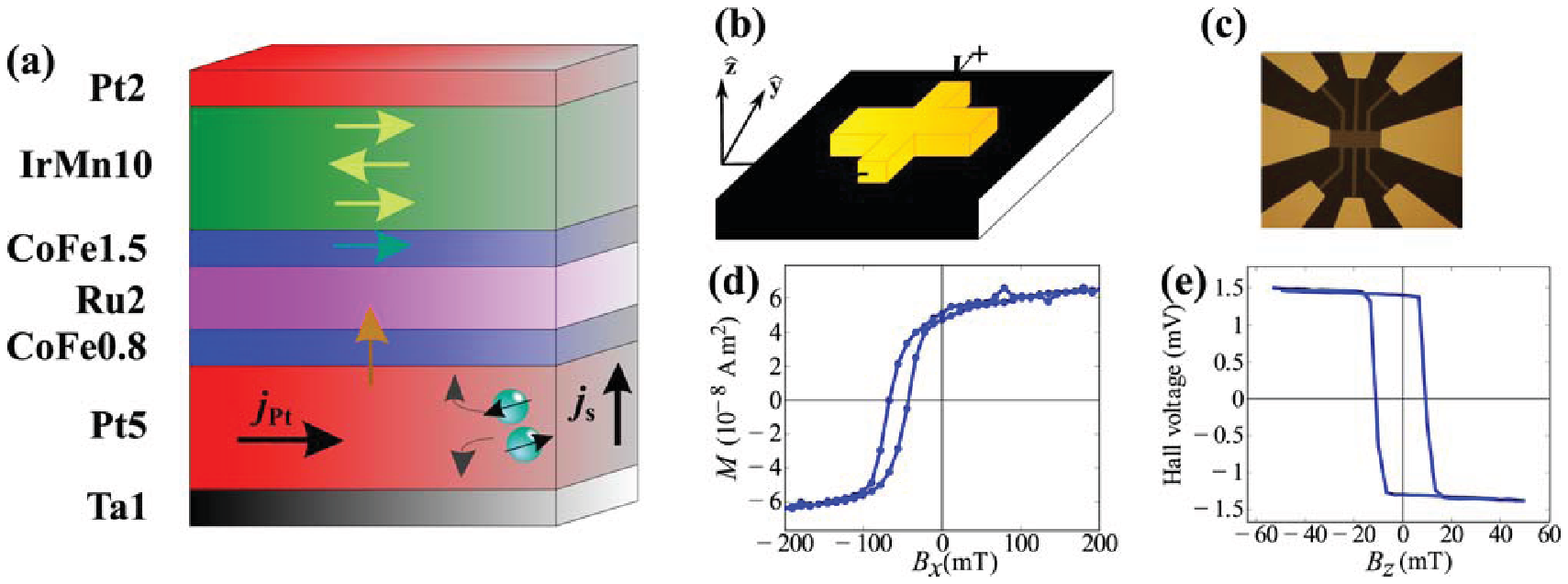}
\caption{(a) Schematic illustration of the stack used for the SOT switching of a
PMA layer. The bottom \ce{CoFe} layer has perpendicular anisotropy, while the
top \ce{CoFe} magnetic moment is pinned along the $\mathbf{x}$ direction by
the
antiferromagnetic \ce{IrMn}. The drawing is not to scale.
(b) Coordinate system and representation of a Hall bar.
(c) Optical microscope image of a device.
(d) In-plane SQUID magnetometry showing the pinned top \ce{CoFe} layer.
(e) Anomalous Hall effect loop showing the bottom free perpendicular \ce{CoFe} layer.
}
\label{fig:sampleschematic}
\end{center}
\end{figure*}

Our stacks, illustrated in Figure\,\ref{fig:sampleschematic}(a), consist of
\ce{Ta}(\num{1})/\ce{Pt}(\num{5})/\ce{CoFe}(\num{0.8})/\ce{Ru}($t_{\ce{Ru}}$)/\ce{CoFe}(\num{1.5})/\ce{IrMn}(\num{10})/\ce{Pt}(\num{2})
(thicknesses in
nanometre).
The \ce{Ru} has been selected for the spacer  because it provides the strongest IEC\cite{Parkin1991} and together with the \ce{Pt} underlayer,
it improves the PMA of the \ce{CoFe} free layer.
A series of control samples with varying \ce{Pt} thickness:
\ce{Si}/\ce{SiO2}(substrate)//\ce{Ta}(\num{1})/\ce{Pt}($t_{\ce{Pt}}$)/\ce{CoFe}(\num{0.8})/\ce{Ru}(\num{3}) were also grown.
All stacks are patterned into micron-sized Hall bars with a channel of
width \SI{20}{\micro\metre} and length ranging from $50$ to \SI{100}{\micro\metre}.
Figure\,\ref{fig:sampleschematic}(b) shows a device schematic with the
definition of the coordinate system, while Figure\,\ref{fig:sampleschematic}(c)
is an optical micrograph of a typical Hall bar. All patterned devices are vacuum
annealed at \SI{250}{\degreeCelsius}
for \SI{1}{\hour} in
\SI{800}{\milli\tesla} to set the exchange-bias direction. The top \ce{CoFe}
layer is pinned along the $\mathbf{x}$-axis, with
the magnetization parallel (anti-parallel) to the current when the
annealing field is
directed along $\mathbf{ x}$ ($-\mathbf{ x}$). An exchange-bias field of
$B_{\textrm{EB}} \sim \SI{50}{\milli\tesla}$ on a blanket film with $t_{\ce{Ru}} =
\SI{2}{\nano\metre}$ is
evidenced by magnetization curve in a field $B_x$, plotted
in Figure\,\ref{fig:sampleschematic}(d).
Figure\,\ref{fig:sampleschematic}(e) shows the anomalous
Hall effect (AHE)
voltage, $V_\textrm{H}$ as a function of out-of-plane field $B_z$ for a Hall bar
with $t_{\ce{Ru}} = \SI{2}{\nano\metre}$. $V_\textrm{H}$ is
measured with an applied current of \SI{2}{\milli\ampere} which corresponds to a
current density of $j_{\ce{Pt}} = \SI {1.5E10}{\ampere\per\square\metre}$ in the
bottom \ce{Pt} layer. 

With the convention defined in Figure\,\ref{fig:sampleschematic}(b),
the spin Hall effect (SHE)\cite{Dyakonov1971,Hirsch1999} in \ce{Pt}
due to a charge current $\mathbf{j}_\ce{Pt}$ along $\mathbf{x}$, 
generates a spin accumulation $\mathbold{\sigma}$
polarised along $- \mathbf{y}$
at the top interface of the \ce{Pt} layer.
The pure spin current, $\mathbf{j}_s$, relaxing within the adjacent \ce{CoFe} free layer with moment $\mathbf{m}$,
exerts a
Slonczewski-like SOT directed along $\mathbf{m} \times (\mathbold{\sigma} \times
\mathbf{m})$ and a
field-like SOT along $\mathbf{m} \times \mathbold{\sigma}$.
The magnitudes of the two orthogonal SOT components are parametrised by the real
and the imaginary parts of the complex
spin-mixing conductance $G_{\uparrow\downarrow} = G' + iG''$ at the
\ce{Pt}/\ce{CoFe}
interface\cite{Brataas.PRL.2000}.
Given the micrometric
dimensions of our devices, a macrospin model is inapplicable and
the switching should
be described in terms of domain nucleation followed by
thermally-assisted SOT-driven domain wall propagation.\cite{Lee.PRB.2014}
Efficient SOT-driven domain wall motion in a
PMA material can be obtained when the wall assumes a N\'{e}el configuration (where the
magnetization rotates in the $xz$ plane) rather than a Bloch one
(where the
magnetization rotates in the $yz$ plane).\cite{Haazen2013,Emori2013}
Bloch walls tend to be
favoured in magnetic structures with PMA where the film
thickness is
negligible compared to other dimensions but an in-plane bias field along
$\mathbf{x}$ of order $B_x \approx \SI{10}{\milli\tesla}$ is sufficient to transform
a Bloch wall into a
N\'{e}el wall\cite{Haazen2013}. Since SOT-driven domain wall motion is opposite for walls of
opposite chirality
(for instance ``$\uparrow\nearrow\rightarrow\searrow\downarrow$" and
``$\downarrow\searrow\rightarrow\nearrow\uparrow$" walls),
 a reversed domain 
will either expand or collapse upon passing a current along the
external field direction\cite{Haazen2013,Emori2013,Lee.PRB.2014,Torrejon2014}.
This leads to deterministic SOT switching where the preferred magnetization
state depends on the sign of the product of the current $\mathbf{j}_\ce{Pt}$ and the $\mathbf{x}$-projection of the domain wall moment $m^\textrm{DW}_x$.
In our device, N\'{e}el domain walls with a particular sign of $m^\textrm{DW}_x$ are stabilised by IEC from top \ce{CoFe}
layer.\cite{Parkin1990}
Robust zero-field switching is achieved by pinning the magnetization of the top \ce{CoFe} layer in
the same direction as the applied current by exchange bias with antiferromagnetic \ce{IrMn}.

\begin{figure*}[ht]
\includegraphics[width=1.0\textwidth]{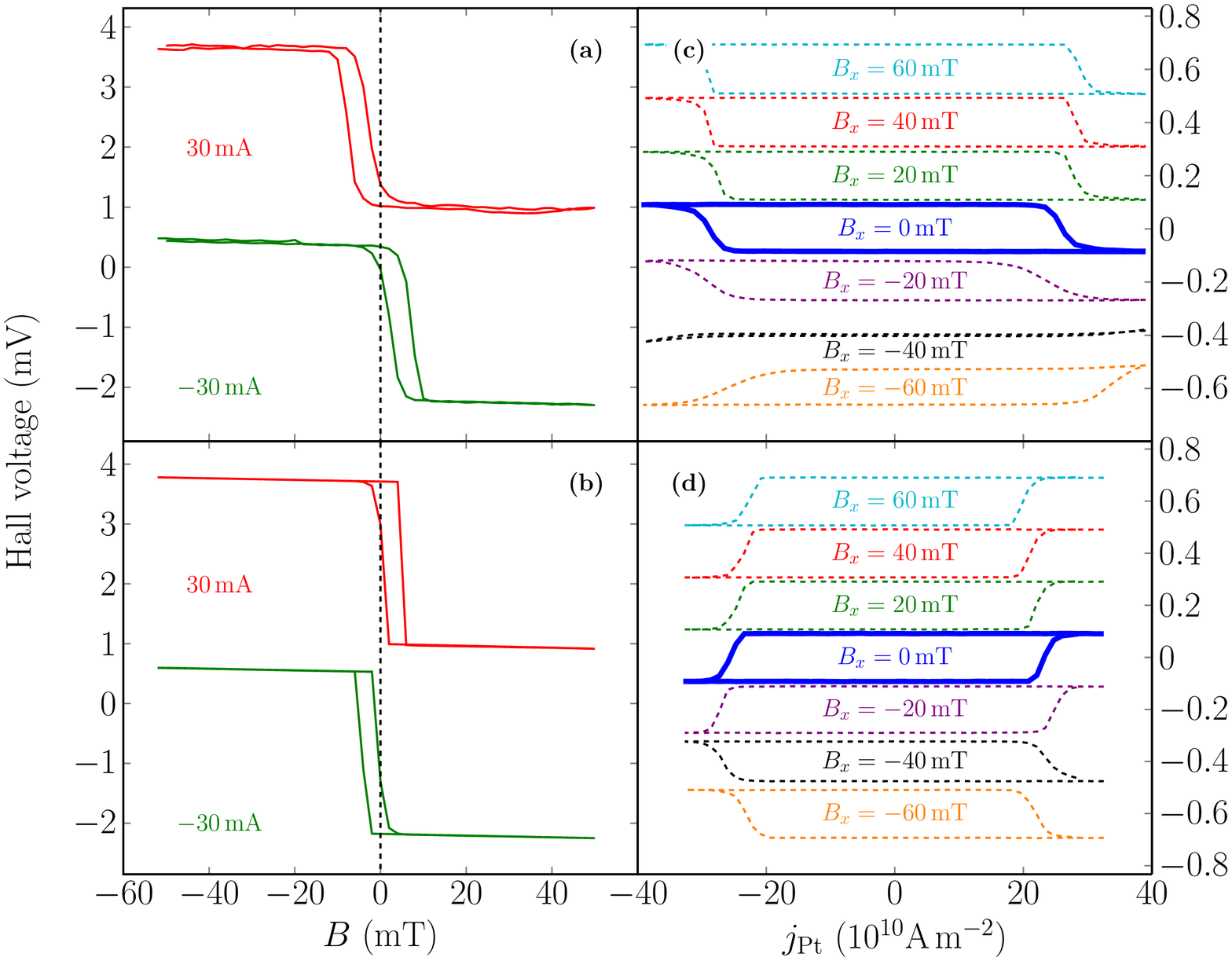}
\caption{(a-b) Anomalous Hall effect voltage as a function of the out-of-plane external field for different
injected curent values. While the low-current loop doesn't show any asymmetry, the high-current loops show a preferred 
orientation of the magnetic moment.
(c-d) Anomalous Hall effect voltage as a function of the injected current
density in the \ce{Pt} layer measured at different external fields $B_x$
along the current direction.  
(a-c) Sample with $t_\ce{Ru} = \SI{2}{\nano\metre}$, showing antiferromagnetic interlayer coupling.
(b-d) Sample with $t_\ce{Ru} = \SI{2.5}{\nano\metre}$, showing ferromagnetic interlayer coupling.
The top \ce{CoFe} layer has been pinned in the $+\mathbf{x}$ direction for both samples.
The loops are shifted for clarity.
}
\label{fig:SOTswitching}
\end{figure*}

We will focus on the switching properties of stacks with
$t_{\ce{Ru}} = \SI{2}{\nano\metre}$  and $t_{\ce{Ru}} = \SI{2.5}{\nano\metre}$ for
which the IEC via \ce{Ru} is respectively antiferromagnetic (AFM) or
ferromagnetic (FM). The pinned layers are exchange-biased along $+\mathbf{ x}$.
Figure\,\ref{fig:SOTswitching}(a-b) shows AHE loops with perpendicular applied field
obtained at $I = \pm \SI{30}{\milli\ampere}$ for the two stacks.
While the loops taken at \SI{2}{\milli\ampere} 
did not show any
noticeable asymmetry in the coercivity for positive and negative fields (Figure\,\ref{fig:sampleschematic}(e)),
there is clearly a preferred switching direction in the
high-current loops. This
is in agreement with the presence of a torque that favours an
orientation of $\mathbf{m}$ that depends on the sign of the
injected current. In addition, upon changing the coupling from AFM to FM, the field
shift of the AHE loops is reversed which indicates a sign change of $m^\textrm{DW}_x$. We further verified that the effect is absent in the
control sample without the \ce{CoFe} pinned layer.

Successive current pulses with a width of \SI{10}{\milli\second} are applied to the device
and the Hall voltage is measured at a lower current of \SI{2}{\milli\ampere}
after each
one to probe the magnetization state of the free layer.
Figure\,\ref{fig:SOTswitching}(c-d) shows the current-induced switching of the PMA
layer of two devices measured at various external applied fields $B_x$.
Both devices exhibit reversible SOT switching in the absence of an
external field, but the
$V_{\textrm{H}}-j_{\ce{Pt}}$ loops of the two devices are opposite in sign due to opposite IEC for the two \ce{Ru} thicknesses.
We interpret this in terms of the presence of N\'{e}el domain walls in the
free layer with a sign of $m^\textrm{DW}_x$, that is stabilised by the magnetic coupling from the pinned
layer mediated via the \ce{Ru} spacer.
We also see that $V_{\textrm{H}}-j_{\ce{Pt}}$ loops are reversed at $B_x \approx -\SI{40}{\milli\tesla}$, which corresponds
to the exchange bias field of the pinned \ce{CoFe}.
The switching of a device therefore depends on the sign of the coupling together with the value of $B_x$ relative to two characteristic fields of the system:
the exchange bias field $B_\textrm{EB}$ and the IEC field $B_\textrm{IEC}$.
The first is the effective field acting on the top \ce{CoFe} layer coming from the direct exchange with the
antiferromagnetic \ce{IrMn},
while the second is the effective field acting on the bottom \ce{CoFe} layer coming from the
oscillatory interlayer coupling with the
top \ce{CoFe} via the \ce{Ru} spacer.
In the absence of external field ($B_x = \SI{0}{\milli\tesla}$), $m^\textrm{DW}_x$ is determined by the magnetization of the pinned layer and the sign of the IEC.
When $B_x$ overcomes $B_\textrm{EB}$ ($-B_x > B_\textrm{EB}$), $\mathbf{m}$
of the pinned \ce{CoFe} is reversed, which consequently changes the orientation
of $m^\textrm{DW}_x$ and flips the $V_\textrm{H}-j_{\ce{Pt}}$ loop. For both devices, the absence of
a sign reversal for $-B_x$ ranging between zero and $B_\textrm{EB}$ indicates that $B_\textrm{IEC} > B_\textrm{EB} \approx \SI{40}{\milli\tesla}$.
Furthermore, one expects to observe the breakdown of the AFM IEC at sufficiently high external bias fields $|B_x| > B_\textrm{IEC}$,
which will again be indicated by the reversal of the $V_\textrm{H}-j_{\ce{Pt}}$ loop. For
$t_{\ce{Ru}} = \SI{2}{\nano\metre}$, the fact that no such reversal is seen for an applied field up
to $|B_x| = \SI{100}{\milli\tesla}$ gives a lower limit to $B_\textrm{IEC}$.
In order to test our model, we also annealed other devices with $t_{\ce{Ru}} = \SI{2}{\nano\metre}$ and $t_{\ce{Ru}} = \SI{2.5}{\nano\metre}$
in a magnetic field $B_x = \SI{-800}{\milli\tesla}$ to set the exchange bias of
the pinned \ce{CoFe} layers in opposite directions compared to devices shown here, hence exerting opposite
coupling to the domain walls within the thin \ce{CoFe}. Those results confirm our explanation.


Finally, we used two independent methods to quantify the SOT in our heterostructures: spin Hall
magnetoresistance
(SMR)\cite{Chen2013,Nakayama.PRL.2013,Althammer.PRB.2013,Avci2015,Kim.Arxiv.2015} and
harmonic Hall measurements\cite{Kim2013,Garello2013}. The measurements were performed on devices
patterned from control stacks without the top pinning layers.

\begin{figure*}[ht]
\includegraphics[
width=1.0\textwidth]{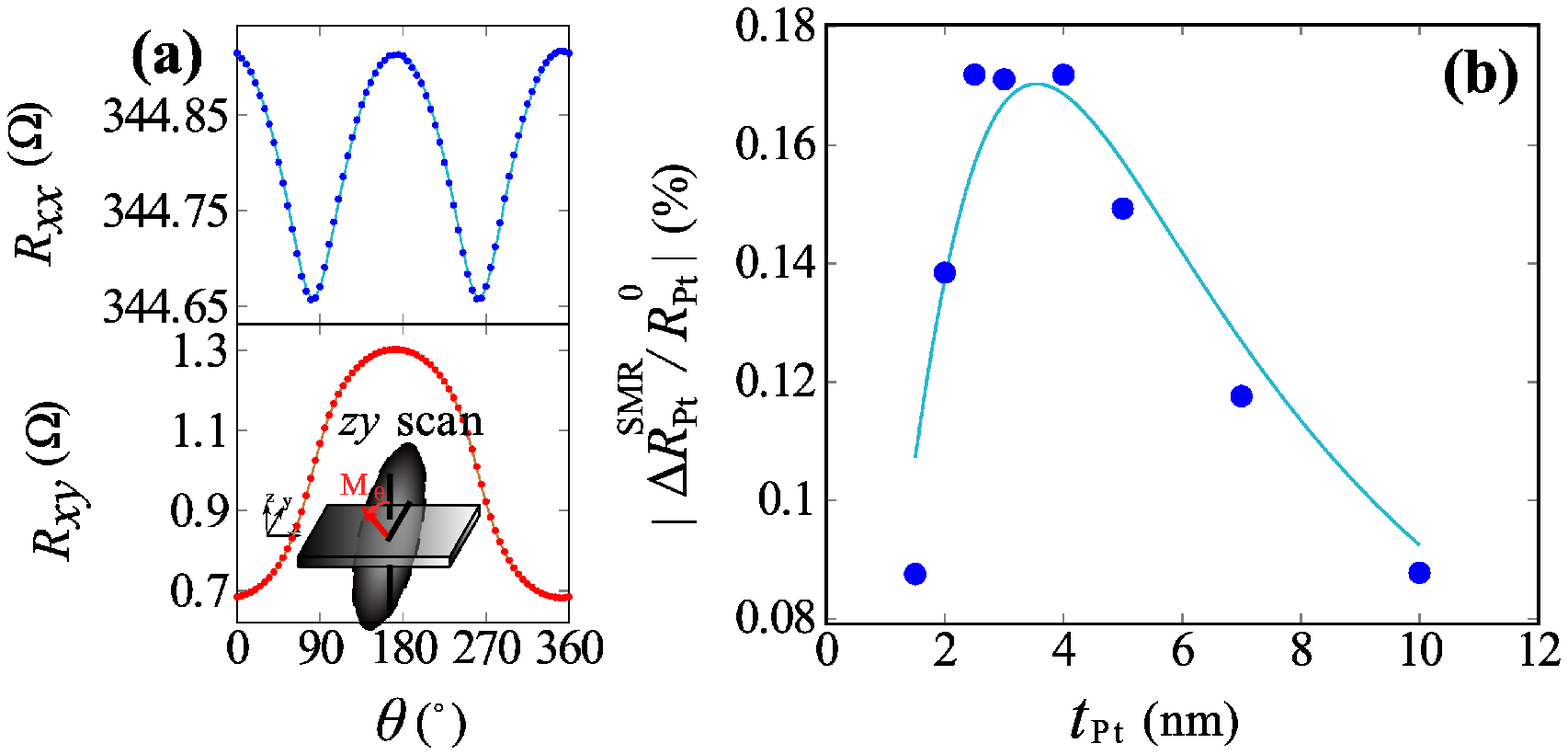}
\caption{Spin Hall magnetoresistance. (a) Longitudinal (upper panel) and transverse (lower panel) resistance for
a sample of \ce{Ta}(\num{1})/\ce{Pt}(\num{2})/\ce{CoFe}(\num{0.8})/\ce{Ru}(\num{3}) as a function
of the magnetization direction in the $zy$ plane.
(b) \ce{Pt} thickness dependence of the spin Hall magnetoresistance
$\Delta R_\ce{Pt}^\textrm{SMR}$. The solid line is the fit using
eq\,\ref{eq:RxxSMRfitequation}. The error bars are given by the size of the dots.
}
\label{fig:smrfitting}
\end{figure*}

The SMR effect (Fig.\,\ref{fig:smrfitting}) is caused by the simultaneous action of
the spin Hall effect and the inverse spin Hall effect due to transmission and
reflection of the spin current at the \ce{Pt}/\ce{CoFe} interface. 
The ratio
between the reflected and the transmitted fractions of the spin accumulation
depends on the relative orientation of the polarization of the
electrons and the magnetic moment of the ferromagnet and on the spin-mixing
conductance $G_{\uparrow\downarrow}$. If the imaginary part $G''$ is negligible compared to the
real part $G'$, the SOT acting on the magnetization has the form $\mathbf{m}
\times (\mathbold{\sigma} \times \mathbf{m})$, with $\mathbold{\sigma} \parallel \mathbf{ y}$. 
We confirm by harmonic Hall measurements that this
is the case for \ce{Pt}\cite{Garello2013, Emori2013}.
The torque is zero for $\mathbf{ m} \parallel \mathbf{ y}$, when the
absorption of spin current is at a minimum and the reflection at a maximum.
Hence the longitudinal resistance $R_{xx}$ of the stack shows a $m_y^2$ dependence due to
SMR of the \ce{Pt}/\ce{CoFe} bilayer.\cite{Nakayama.PRL.2013,Althammer.PRB.2013,Avci2015,Kim.Arxiv.2015}
In addition, the effect of the SMR is seen in the transverse resistance
$R_{xy}$ as an additional contribution to the planar Hall effect with an $m_xm_y$ dependency on the
magnetization orientation.
Angular scans of $R_{xx}$ and $R_{xy}$ in the $zy$ plane are shown in Fig.\,\ref{fig:smrfitting}(a).
Assuming \ce{Pt} is the unique source of SMR, it would be more 
appropriate to use the magnetoresistance due to SMR \textit{within} the \ce{Pt} layer 
$\Delta R_{\ce{Pt}}^\textrm{SMR}$ instead of the measured overall $\Delta R_{xx}^\textrm{SMR}$
for rigorous SMR analysis
$\Delta R_{\ce{Pt}}^\textrm{SMR}$ is related to $\theta_\textrm{SH}$, the spin Hall
angle of the $\ce{Pt}/\ce{CoFe}$ system by the equation\cite{Althammer.PRB.2013,Kim.Arxiv.2015}
\begin{equation}
\label{eq:RxxSMRfitequation}
 \frac{\Delta R_{\ce{Pt}}^\textrm{SMR}}{R_{Pt}^0} = -
\frac{\theta_\textrm{SH}^2}{t_{\ce{Pt}}} \frac{2 \lambda_{\textrm{sf}}^2
\rho_{\ce{Pt}} G'\tanh^2\frac{t_{\ce{Pt}}}{2\lambda_{\textrm{sf}}}}{1 + 2
\lambda_{\textrm{sf}}
\rho_{\ce{Pt}} G'\coth\frac{t_{\ce{Pt}}}{\lambda_{\textrm{sf}}}}
\end{equation}
where $R_{\ce{Pt}}^0$ is the resistance of the \ce{Pt} underlayer without SMR
contributions, $\rho_{\ce{Pt}}$ is its resistivity and
$\lambda_\textrm{sf}$ is its spin diffusion length.
The values of $\Delta R_{\ce{Pt}}^\textrm{SMR}$ have been determined by measuring
the variation in the longitudinal resistance while rotating a \SI{2}{\tesla} external
magnetic field in the $zy$ plane.
The best fit of the \ce{Pt}
thickness $t_{\ce{Pt}}$ dependence of $\Delta R_{\ce{Pt}}^\textrm{SMR}/R^0_{\ce{Pt}}$
with $\theta_\textrm{SH}$, $G'$ and $\lambda_\textrm{sf}$ as parameters
is shown in Figure\,\ref{fig:smrfitting}(b).
Using $\rho_{\ce{Pt}} = \SI{26}{\micro\ohm\centi\metre}$, we obtained
$\lambda_\textrm{sf} = \SI{1.6}{\nano\metre}$, $G'
=\SI{7.5e14}{\per\ohm\per\metre}$ and $\theta_\textrm{SH}
= \SI{12.4}{\percent}$, all in excellent agreement with a previous report
on the \ce{YIG}/\ce{Pt} system.\cite{Althammer.PRB.2013}

We also measured the first and second harmonic
Hall responses of the same device with $t_{\ce{Pt}} = \SI{5}{\nano\metre}$ under a low
frequency ac current excitation to quantify the effective spin-orbit fields.\cite{Kim2013,Garello2013} Considering the
Hall voltage contribution from the anisotropic magnetoresistance and the SMR, we derive the longitudinal and transverse effective spin-orbit fields.
As previously mentioned, the transverse field is negligible for \ce{Pt}, confirming that the imaginary part of the spin-mixing
conductance and the Rashba effect are very small. We found a spin Hall
angle of $\theta_\textrm{SH} = \SI{11.2}{\percent}$, in good agreement with our SMR analysis.

In summary, we have demonstrated a novel approach to achieve zero-field SOT
switching using the IEC via
a nonmagnetic spacer. The preferred magnetization
state of the free layer with PMA is reversed upon reversing the
current polarity, the exchange-bias direction or the exchange
coupling sign.
The coupling cannot be explained by the stray field from a flat top \ce{CoFe} layer, which
is only antiferromagnetic and very small ($\sim\SI{0.001}{\milli\tesla}$),
nor by the stray field created
by correlated surface roughness because the N\'{e}el orange peel
mechanism\cite{Neel1962} is always ferromagnetic.

Our approach is scalable
because the mechanism is independent of the area of the device as long as the dimensions 
are greater than the domain wall width, which is of the order of \SI{25}{\nano\metre}.
Furthermore, SOT switching has been demonstrated down to \SI{30}{\nano\metre} dots
with an applied field of \SI{20}{\milli\tesla}\cite{Zhang2015}, which is well within the capability of our structure.

It should be possible to develop a new 
three-terminal device with magnetoresistive readout by using a
SOT layer which is also the spacer for magnetic coupling. An iridium spacer
which exhibits high spin-orbit coupling and relatively strong IEC\cite{Parkin1991}
might work.
Our new switching concept, which is based on well-understood phenomena and materials,
takes us a step closer to the practical realisation of spin-orbit torque
applications involving manipulation of perpendicular
nanomagnets, which include SOT-MRAM, SOT-based magnetic logic and an
SOT-based magnetic racetrack.

\section*{Methods}
\label{sec:methods}
\subsection*{Sample and device fabrication}
The stacks for demonstrating zero-field SOT switching are, from the substrate,
\ce{Ta}(\num{1})/\ce{Pt}(\num{5})/\ce{CoFe}(\num{0.8})/\ce{Ru}($t_{\ce{Ru}}$)/\ce{CoFe}(\num{1.5})/\ce{IrMn}(\num{10})/\ce{Pt}(\num{2})
(thicknesses in nanometres)
and those for SMR measurements and \ce{Pt} resistivity fitting are \ce{Ta}(\num{1})/\ce{Pt}($t_{\ce{Pt}}$)/\ce{CoFe}(\num{0.8})/\ce{Ru}(\num{3}) with 
$t_{\ce{Pt}}$ ranging from \SI{1}{\nano\metre} to \SI{10}{\nano\metre}.
The $t_{\ce{Pt}} = \SI{5}{\nano\metre}$ sample from the latter series is also used for the harmonic Hall measurement
and served as the reference for the switching experiment.
All stacks are deposited on \ce{Si}(001) substrates with \SI{500}{\nano\metre} thermal oxide.
Layers are grown by d.c. magnetron sputtering using an automated Shamrock sputtering tool with a
chamber base pressure of \SI{3E-7}{\torr} and a growth pressure of $\approx \SI{2}{\milli\torr}$. The growth rates of various metals are lower than
\SI{0.03}{\nano\metre\per\second}, calibrated using X-ray reflectometry. The bottom \ce{Ta}(\num{1})
layer serves as an adhesion layer for the \ce{Pt}, improving the PMA of the \ce{CoFe}. 
Eight-terminal Hall bar devices are fabricated using standard ultra-violet optical lithography and \ce{Ar} ion milling.
The \ce{Ti}(\num{5})/\ce{Cu}(\num{80})/\ce{Au}(\num{20}) contacts are formed by electron-beam evaporation. 

\subsection*{Characterisation}
All sample characterisation is performed at room temperature.
The saturation magnetization of the thin \ce{CoFe} layer with PMA, obtained from the SQUID magnetometry, is
$\approx \SI{1.2}{\mega\ampere\per\metre}$, a value lower than the bulk.
The angular dependence of the longitudinal and transverse magnetoresistance is measured at a d.c. current of \SI{2}{\milli\ampere}
by fixing the device in a rotating \SI{2.0}{\tesla} magnetic field produced by a Multimag 
permanent magnet variable flux source. In harmonic Hall measurements,
sinusoidal a.c. excitation is generated by a WF1946B waveform generator at a frequency of \SI{1234.57}{\hertz} and the current 
is measured on a \SI{100}{\ohm} series resistor using a EG\&G 5210 lock-in amplifier.
A small in-plane field is swept while
the in-phase first harmonic and the out-of-phase second harmonic Hall signals are
simultaneously detected using two SR830 lock-in amplifiers. 

\bibliography{BettoDavide_arxiv2}

\begin{thebibliography}{30}%
\makeatletter
\providecommand \@ifxundefined [1]{%
 \@ifx{#1\undefined}
}%
\providecommand \@ifnum [1]{%
 \ifnum #1\expandafter \@firstoftwo
 \else \expandafter \@secondoftwo
 \fi
}%
\providecommand \@ifx [1]{%
 \ifx #1\expandafter \@firstoftwo
 \else \expandafter \@secondoftwo
 \fi
}%
\providecommand \natexlab [1]{#1}%
\providecommand \enquote  [1]{``#1''}%
\providecommand \bibnamefont  [1]{#1}%
\providecommand \bibfnamefont [1]{#1}%
\providecommand \citenamefont [1]{#1}%
\providecommand \href@noop [0]{\@secondoftwo}%
\providecommand \href [0]{\begingroup \@sanitize@url \@href}%
\providecommand \@href[1]{\@@startlink{#1}\@@href}%
\providecommand \@@href[1]{\endgroup#1\@@endlink}%
\providecommand \@sanitize@url [0]{\catcode `\\12\catcode `\$12\catcode
  `\&12\catcode `\#12\catcode `\^12\catcode `\_12\catcode `\%12\relax}%
\providecommand \@@startlink[1]{}%
\providecommand \@@endlink[0]{}%
\providecommand \url  [0]{\begingroup\@sanitize@url \@url }%
\providecommand \@url [1]{\endgroup\@href {#1}{\urlprefix }}%
\providecommand \urlprefix  [0]{URL }%
\providecommand \Eprint [0]{\href }%
\providecommand \doibase [0]{http://dx.doi.org/}%
\providecommand \selectlanguage [0]{\@gobble}%
\providecommand \bibinfo  [0]{\@secondoftwo}%
\providecommand \bibfield  [0]{\@secondoftwo}%
\providecommand \translation [1]{[#1]}%
\providecommand \BibitemOpen [0]{}%
\providecommand \bibitemStop [0]{}%
\providecommand \bibitemNoStop [0]{.\EOS\space}%
\providecommand \EOS [0]{\spacefactor3000\relax}%
\providecommand \BibitemShut  [1]{\csname bibitem#1\endcsname}%
\let\auto@bib@innerbib\@empty
\bibitem [{\citenamefont {Miron}\ \emph {et~al.}(2011)\citenamefont {Miron},
  \citenamefont {Garello}, \citenamefont {Gaudin}, \citenamefont {Zermatten},
  \citenamefont {Costache}, \citenamefont {Auffret}, \citenamefont {Bandiera},
  \citenamefont {Rodmacq}, \citenamefont {Schuhl},\ and\ \citenamefont
  {Gambardella}}]{Miron2011}%
  \BibitemOpen
  \bibfield  {author} {\bibinfo {author} {\bibfnamefont {I.~M.}\ \bibnamefont
  {Miron}}, \bibinfo {author} {\bibfnamefont {K.}~\bibnamefont {Garello}},
  \bibinfo {author} {\bibfnamefont {G.}~\bibnamefont {Gaudin}}, \bibinfo
  {author} {\bibfnamefont {P.-J.}\ \bibnamefont {Zermatten}}, \bibinfo {author}
  {\bibfnamefont {M.~V.}\ \bibnamefont {Costache}}, \bibinfo {author}
  {\bibfnamefont {S.}~\bibnamefont {Auffret}}, \bibinfo {author} {\bibfnamefont
  {S.}~\bibnamefont {Bandiera}}, \bibinfo {author} {\bibfnamefont
  {B.}~\bibnamefont {Rodmacq}}, \bibinfo {author} {\bibfnamefont
  {A.}~\bibnamefont {Schuhl}}, \ and\ \bibinfo {author} {\bibfnamefont
  {P.}~\bibnamefont {Gambardella}},\ }\href
  {http://dx.doi.org/10.1038/nature10309} {\bibfield  {journal} {\bibinfo
  {journal} {Nature}\ }\textbf {\bibinfo {volume} {476}},\ \bibinfo {pages}
  {189} (\bibinfo {year} {2011})}\BibitemShut {NoStop}%
\bibitem [{\citenamefont {Liu}\ \emph {et~al.}(2012{\natexlab{a}})\citenamefont
  {Liu}, \citenamefont {Lee}, \citenamefont {Gudmundsen}, \citenamefont
  {Ralph},\ and\ \citenamefont {Buhrman}}]{Liu.PRL.2012a}%
  \BibitemOpen
  \bibfield  {author} {\bibinfo {author} {\bibfnamefont {L.}~\bibnamefont
  {Liu}}, \bibinfo {author} {\bibfnamefont {O.~J.}\ \bibnamefont {Lee}},
  \bibinfo {author} {\bibfnamefont {T.~J.}\ \bibnamefont {Gudmundsen}},
  \bibinfo {author} {\bibfnamefont {D.~C.}\ \bibnamefont {Ralph}}, \ and\
  \bibinfo {author} {\bibfnamefont {R.~A.}\ \bibnamefont {Buhrman}},\ }\href
  {\doibase 10.1103/PhysRevLett.109.096602} {\bibfield  {journal} {\bibinfo
  {journal} {Phys. Rev. Lett.}\ }\textbf {\bibinfo {volume} {109}},\ \bibinfo
  {pages} {096602} (\bibinfo {year} {2012}{\natexlab{a}})}\BibitemShut
  {NoStop}%
\bibitem [{\citenamefont {Liu}\ \emph {et~al.}(2012{\natexlab{b}})\citenamefont
  {Liu}, \citenamefont {Pai}, \citenamefont {Li}, \citenamefont {Tseng},
  \citenamefont {Ralph},\ and\ \citenamefont {Buhrman}}]{Liu.Science.2012}%
  \BibitemOpen
  \bibfield  {author} {\bibinfo {author} {\bibfnamefont {L.}~\bibnamefont
  {Liu}}, \bibinfo {author} {\bibfnamefont {C.-F.}\ \bibnamefont {Pai}},
  \bibinfo {author} {\bibfnamefont {Y.}~\bibnamefont {Li}}, \bibinfo {author}
  {\bibfnamefont {H.~W.}\ \bibnamefont {Tseng}}, \bibinfo {author}
  {\bibfnamefont {D.~C.}\ \bibnamefont {Ralph}}, \ and\ \bibinfo {author}
  {\bibfnamefont {R.~A.}\ \bibnamefont {Buhrman}},\ }\href {\doibase
  10.1126/science.1218197} {\bibfield  {journal} {\bibinfo  {journal}
  {Science}\ }\textbf {\bibinfo {volume} {336}},\ \bibinfo {pages} {555}
  (\bibinfo {year} {2012}{\natexlab{b}})},\ \Eprint
  {http://arxiv.org/abs/http://www.sciencemag.org/content/336/6081/555.full.pdf}
  {http://www.sciencemag.org/content/336/6081/555.full.pdf} \BibitemShut
  {NoStop}%
\bibitem [{\citenamefont {Fan}\ \emph {et~al.}(2014)\citenamefont {Fan},
  \citenamefont {Upadhyaya}, \citenamefont {Kou}, \citenamefont {Lang},
  \citenamefont {Takei}, \citenamefont {Wang}, \citenamefont {Tang},
  \citenamefont {He}, \citenamefont {Chang}, \citenamefont {Montazeri},
  \citenamefont {Yu}, \citenamefont {Jiang}, \citenamefont {Nie}, \citenamefont
  {Schwartz}, \citenamefont {Tserkovnyak},\ and\ \citenamefont
  {Wang}}]{Fan.NatMat.2014}%
  \BibitemOpen
  \bibfield  {author} {\bibinfo {author} {\bibfnamefont {Y.}~\bibnamefont
  {Fan}}, \bibinfo {author} {\bibfnamefont {P.}~\bibnamefont {Upadhyaya}},
  \bibinfo {author} {\bibfnamefont {X.}~\bibnamefont {Kou}}, \bibinfo {author}
  {\bibfnamefont {M.}~\bibnamefont {Lang}}, \bibinfo {author} {\bibfnamefont
  {S.}~\bibnamefont {Takei}}, \bibinfo {author} {\bibfnamefont
  {Z.}~\bibnamefont {Wang}}, \bibinfo {author} {\bibfnamefont {J.}~\bibnamefont
  {Tang}}, \bibinfo {author} {\bibfnamefont {L.}~\bibnamefont {He}}, \bibinfo
  {author} {\bibfnamefont {L.-T.}\ \bibnamefont {Chang}}, \bibinfo {author}
  {\bibfnamefont {M.}~\bibnamefont {Montazeri}}, \bibinfo {author}
  {\bibfnamefont {G.}~\bibnamefont {Yu}}, \bibinfo {author} {\bibfnamefont
  {W.}~\bibnamefont {Jiang}}, \bibinfo {author} {\bibfnamefont
  {T.}~\bibnamefont {Nie}}, \bibinfo {author} {\bibfnamefont {R.~N.}\
  \bibnamefont {Schwartz}}, \bibinfo {author} {\bibfnamefont {Y.}~\bibnamefont
  {Tserkovnyak}}, \ and\ \bibinfo {author} {\bibfnamefont {K.~L.}\ \bibnamefont
  {Wang}},\ }\href {http://dx.doi.org/10.1038/nmat3973} {\bibfield  {journal}
  {\bibinfo  {journal} {Nat Mater}\ }\textbf {\bibinfo {volume} {13}},\
  \bibinfo {pages} {699} (\bibinfo {year} {2014})}\BibitemShut {NoStop}%
\bibitem [{\citenamefont {Slonczewski}(1996)}]{Slonczewski1996}%
  \BibitemOpen
  \bibfield  {author} {\bibinfo {author} {\bibfnamefont {J.}~\bibnamefont
  {Slonczewski}},\ }\href {\doibase
  http://dx.doi.org/10.1016/0304-8853(96)00062-5} {\bibfield  {journal}
  {\bibinfo  {journal} {Journal of Magnetism and Magnetic Materials}\ }\textbf
  {\bibinfo {volume} {159}},\ \bibinfo {pages} {L1 } (\bibinfo {year}
  {1996})}\BibitemShut {NoStop}%
\bibitem [{\citenamefont {Berger}(1996)}]{Berger.PRB.1996}%
  \BibitemOpen
  \bibfield  {author} {\bibinfo {author} {\bibfnamefont {L.}~\bibnamefont
  {Berger}},\ }\href {\doibase 10.1103/PhysRevB.54.9353} {\bibfield  {journal}
  {\bibinfo  {journal} {Phys. Rev. B}\ }\textbf {\bibinfo {volume} {54}},\
  \bibinfo {pages} {9353} (\bibinfo {year} {1996})}\BibitemShut {NoStop}%
\bibitem [{\citenamefont {Ikeda}\ \emph {et~al.}(2010)\citenamefont {Ikeda},
  \citenamefont {Miura}, \citenamefont {Yamamoto}, \citenamefont {Mizunuma},
  \citenamefont {Gan}, \citenamefont {Endo}, \citenamefont {Kanai},
  \citenamefont {Hayakawa}, \citenamefont {Matsukura},\ and\ \citenamefont
  {Ohno}}]{Ikeda2010}%
  \BibitemOpen
  \bibfield  {author} {\bibinfo {author} {\bibfnamefont {S.}~\bibnamefont
  {Ikeda}}, \bibinfo {author} {\bibfnamefont {K.}~\bibnamefont {Miura}},
  \bibinfo {author} {\bibfnamefont {H.}~\bibnamefont {Yamamoto}}, \bibinfo
  {author} {\bibfnamefont {K.}~\bibnamefont {Mizunuma}}, \bibinfo {author}
  {\bibfnamefont {H.~D.}\ \bibnamefont {Gan}}, \bibinfo {author} {\bibfnamefont
  {M.}~\bibnamefont {Endo}}, \bibinfo {author} {\bibfnamefont {S.}~\bibnamefont
  {Kanai}}, \bibinfo {author} {\bibfnamefont {J.}~\bibnamefont {Hayakawa}},
  \bibinfo {author} {\bibfnamefont {F.}~\bibnamefont {Matsukura}}, \ and\
  \bibinfo {author} {\bibfnamefont {H.}~\bibnamefont {Ohno}},\ }\href
  {http://dx.doi.org/10.1038/nmat2804} {\bibfield  {journal} {\bibinfo
  {journal} {Nat Mater}\ }\textbf {\bibinfo {volume} {9}},\ \bibinfo {pages}
  {721} (\bibinfo {year} {2010})}\BibitemShut {NoStop}%
\bibitem [{\citenamefont {Parkin}\ \emph {et~al.}(1990)\citenamefont {Parkin},
  \citenamefont {More},\ and\ \citenamefont {Roche}}]{Parkin1990}%
  \BibitemOpen
  \bibfield  {author} {\bibinfo {author} {\bibfnamefont {S.~S.~P.}\
  \bibnamefont {Parkin}}, \bibinfo {author} {\bibfnamefont {N.}~\bibnamefont
  {More}}, \ and\ \bibinfo {author} {\bibfnamefont {K.~P.}\ \bibnamefont
  {Roche}},\ }\href {\doibase 10.1103/PhysRevLett.64.2304} {\bibfield
  {journal} {\bibinfo  {journal} {Phys. Rev. Lett.}\ }\textbf {\bibinfo
  {volume} {64}},\ \bibinfo {pages} {2304} (\bibinfo {year}
  {1990})}\BibitemShut {NoStop}%
\bibitem [{\citenamefont {Mangin}\ \emph {et~al.}(2006)\citenamefont {Mangin},
  \citenamefont {Ravelosona}, \citenamefont {Katine}, \citenamefont {Carey},
  \citenamefont {Terris},\ and\ \citenamefont {Fullerton}}]{Mangin2006}%
  \BibitemOpen
  \bibfield  {author} {\bibinfo {author} {\bibfnamefont {S.}~\bibnamefont
  {Mangin}}, \bibinfo {author} {\bibfnamefont {D.}~\bibnamefont {Ravelosona}},
  \bibinfo {author} {\bibfnamefont {J.~A.}\ \bibnamefont {Katine}}, \bibinfo
  {author} {\bibfnamefont {M.~J.}\ \bibnamefont {Carey}}, \bibinfo {author}
  {\bibfnamefont {B.~D.}\ \bibnamefont {Terris}}, \ and\ \bibinfo {author}
  {\bibfnamefont {E.~E.}\ \bibnamefont {Fullerton}},\ }\href
  {http://dx.doi.org/10.1038/nmat1595} {\bibfield  {journal} {\bibinfo
  {journal} {Nat Mater}\ }\textbf {\bibinfo {volume} {5}},\ \bibinfo {pages}
  {210} (\bibinfo {year} {2006})}\BibitemShut {NoStop}%
\bibitem [{\citenamefont {Yu}\ \emph {et~al.}(2014)\citenamefont {Yu},
  \citenamefont {Upadhyaya}, \citenamefont {Fan}, \citenamefont {Alzate},
  \citenamefont {Jiang}, \citenamefont {Wong}, \citenamefont {Takei},
  \citenamefont {Bender}, \citenamefont {Chang}, \citenamefont {Jiang},
  \citenamefont {Lang}, \citenamefont {Tang}, \citenamefont {Wang},
  \citenamefont {Tserkovnyak}, \citenamefont {Amiri},\ and\ \citenamefont
  {Wang}}]{Yu2014}%
  \BibitemOpen
  \bibfield  {author} {\bibinfo {author} {\bibfnamefont {G.}~\bibnamefont
  {Yu}}, \bibinfo {author} {\bibfnamefont {P.}~\bibnamefont {Upadhyaya}},
  \bibinfo {author} {\bibfnamefont {Y.}~\bibnamefont {Fan}}, \bibinfo {author}
  {\bibfnamefont {J.~G.}\ \bibnamefont {Alzate}}, \bibinfo {author}
  {\bibfnamefont {W.}~\bibnamefont {Jiang}}, \bibinfo {author} {\bibfnamefont
  {K.~L.}\ \bibnamefont {Wong}}, \bibinfo {author} {\bibfnamefont
  {S.}~\bibnamefont {Takei}}, \bibinfo {author} {\bibfnamefont {S.~A.}\
  \bibnamefont {Bender}}, \bibinfo {author} {\bibfnamefont {L.-T.}\
  \bibnamefont {Chang}}, \bibinfo {author} {\bibfnamefont {Y.}~\bibnamefont
  {Jiang}}, \bibinfo {author} {\bibfnamefont {M.}~\bibnamefont {Lang}},
  \bibinfo {author} {\bibfnamefont {J.}~\bibnamefont {Tang}}, \bibinfo {author}
  {\bibfnamefont {Y.}~\bibnamefont {Wang}}, \bibinfo {author} {\bibfnamefont
  {Y.}~\bibnamefont {Tserkovnyak}}, \bibinfo {author} {\bibfnamefont {P.~K.}\
  \bibnamefont {Amiri}}, \ and\ \bibinfo {author} {\bibfnamefont {K.~L.}\
  \bibnamefont {Wang}},\ }\href {http://dx.doi.org/10.1038/nnano.2014.94}
  {\bibfield  {journal} {\bibinfo  {journal} {Nat Nano}\ }\textbf {\bibinfo
  {volume} {9}},\ \bibinfo {pages} {548} (\bibinfo {year} {2014})}\BibitemShut
  {NoStop}%
\bibitem [{\citenamefont {You}\ \emph {et~al.}()\citenamefont {You},
  \citenamefont {Lee}, \citenamefont {Bhowmik}, \citenamefont {Labanowski},
  \citenamefont {Hong}, \citenamefont {Bokor},\ and\ \citenamefont
  {Salahuddin}}]{You.Arxiv.2014}%
  \BibitemOpen
  \bibfield  {author} {\bibinfo {author} {\bibfnamefont {L.}~\bibnamefont
  {You}}, \bibinfo {author} {\bibfnamefont {O.}~\bibnamefont {Lee}}, \bibinfo
  {author} {\bibfnamefont {D.}~\bibnamefont {Bhowmik}}, \bibinfo {author}
  {\bibfnamefont {D.}~\bibnamefont {Labanowski}}, \bibinfo {author}
  {\bibfnamefont {J.}~\bibnamefont {Hong}}, \bibinfo {author} {\bibfnamefont
  {J.}~\bibnamefont {Bokor}}, \ and\ \bibinfo {author} {\bibfnamefont
  {S.}~\bibnamefont {Salahuddin}},\ }\href@noop {} {\enquote {\bibinfo {title}
  {Switching of perpendicularly polarized nanomagnets with spin orbit torque
  without an external magnetic field by engineering a tilted anisotropy},}\
  }\Eprint {http://arxiv.org/abs/arXiv:1409.0620} {arXiv:1409.0620}
  \BibitemShut {NoStop}%
\bibitem [{not()}]{note_on_Fukami}%
  \BibitemOpen
  \href@noop {} {}\bibinfo {note} {SOT switching without external field, based
  on a antiferromagnet/ferromagnet bilayer has been recently reported by Fukami
  et al. in arXiv:1507.00888.}\BibitemShut {Stop}%
\bibitem [{\citenamefont {Nogu\'{e}s}\ and\ \citenamefont
  {Schuller}(1999)}]{Nogues1999}%
  \BibitemOpen
  \bibfield  {author} {\bibinfo {author} {\bibfnamefont {J.}~\bibnamefont
  {Nogu\'{e}s}}\ and\ \bibinfo {author} {\bibfnamefont {I.~K.}\ \bibnamefont
  {Schuller}},\ }\href {\doibase
  http://dx.doi.org/10.1016/S0304-8853(98)00266-2} {\bibfield  {journal}
  {\bibinfo  {journal} {Journal of Magnetism and Magnetic Materials}\ }\textbf
  {\bibinfo {volume} {192}},\ \bibinfo {pages} {203 } (\bibinfo {year}
  {1999})}\BibitemShut {NoStop}%
\bibitem [{\citenamefont {Parkin}(1991)}]{Parkin1991}%
  \BibitemOpen
  \bibfield  {author} {\bibinfo {author} {\bibfnamefont {S.~S.~P.}\
  \bibnamefont {Parkin}},\ }\href {\doibase 10.1103/PhysRevLett.67.3598}
  {\bibfield  {journal} {\bibinfo  {journal} {Phys. Rev. Lett.}\ }\textbf
  {\bibinfo {volume} {67}},\ \bibinfo {pages} {3598} (\bibinfo {year}
  {1991})}\BibitemShut {NoStop}%
\bibitem [{\citenamefont {Dyakonov}\ and\ \citenamefont
  {Perel}(1971)}]{Dyakonov1971}%
  \BibitemOpen
  \bibfield  {author} {\bibinfo {author} {\bibfnamefont {M.}~\bibnamefont
  {Dyakonov}}\ and\ \bibinfo {author} {\bibfnamefont {V.}~\bibnamefont
  {Perel}},\ }\href {\doibase http://dx.doi.org/10.1016/0375-9601(71)90196-4}
  {\bibfield  {journal} {\bibinfo  {journal} {Physics Letters A}\ }\textbf
  {\bibinfo {volume} {35}},\ \bibinfo {pages} {459 } (\bibinfo {year}
  {1971})}\BibitemShut {NoStop}%
\bibitem [{\citenamefont {Hirsch}(1999)}]{Hirsch1999}%
  \BibitemOpen
  \bibfield  {author} {\bibinfo {author} {\bibfnamefont {J.~E.}\ \bibnamefont
  {Hirsch}},\ }\href {\doibase 10.1103/PhysRevLett.83.1834} {\bibfield
  {journal} {\bibinfo  {journal} {Phys. Rev. Lett.}\ }\textbf {\bibinfo
  {volume} {83}},\ \bibinfo {pages} {1834} (\bibinfo {year}
  {1999})}\BibitemShut {NoStop}%
\bibitem [{\citenamefont {Brataas}\ \emph {et~al.}(2000)\citenamefont
  {Brataas}, \citenamefont {Nazarov},\ and\ \citenamefont
  {Bauer}}]{Brataas.PRL.2000}%
  \BibitemOpen
  \bibfield  {author} {\bibinfo {author} {\bibfnamefont {A.}~\bibnamefont
  {Brataas}}, \bibinfo {author} {\bibfnamefont {Y.~V.}\ \bibnamefont
  {Nazarov}}, \ and\ \bibinfo {author} {\bibfnamefont {G.~E.~W.}\ \bibnamefont
  {Bauer}},\ }\href {\doibase 10.1103/PhysRevLett.84.2481} {\bibfield
  {journal} {\bibinfo  {journal} {Phys. Rev. Lett.}\ }\textbf {\bibinfo
  {volume} {84}},\ \bibinfo {pages} {2481} (\bibinfo {year}
  {2000})}\BibitemShut {NoStop}%
\bibitem [{\citenamefont {Lee}\ \emph {et~al.}(2014)\citenamefont {Lee},
  \citenamefont {Liu}, \citenamefont {Pai}, \citenamefont {Li}, \citenamefont
  {Tseng}, \citenamefont {Gowtham}, \citenamefont {Park}, \citenamefont
  {Ralph},\ and\ \citenamefont {Buhrman}}]{Lee.PRB.2014}%
  \BibitemOpen
  \bibfield  {author} {\bibinfo {author} {\bibfnamefont {O.~J.}\ \bibnamefont
  {Lee}}, \bibinfo {author} {\bibfnamefont {L.~Q.}\ \bibnamefont {Liu}},
  \bibinfo {author} {\bibfnamefont {C.~F.}\ \bibnamefont {Pai}}, \bibinfo
  {author} {\bibfnamefont {Y.}~\bibnamefont {Li}}, \bibinfo {author}
  {\bibfnamefont {H.~W.}\ \bibnamefont {Tseng}}, \bibinfo {author}
  {\bibfnamefont {P.~G.}\ \bibnamefont {Gowtham}}, \bibinfo {author}
  {\bibfnamefont {J.~P.}\ \bibnamefont {Park}}, \bibinfo {author}
  {\bibfnamefont {D.~C.}\ \bibnamefont {Ralph}}, \ and\ \bibinfo {author}
  {\bibfnamefont {R.~A.}\ \bibnamefont {Buhrman}},\ }\href {\doibase
  10.1103/PhysRevB.89.024418} {\bibfield  {journal} {\bibinfo  {journal} {Phys.
  Rev. B}\ }\textbf {\bibinfo {volume} {89}},\ \bibinfo {pages} {024418}
  (\bibinfo {year} {2014})}\BibitemShut {NoStop}%
\bibitem [{\citenamefont {Haazen}\ \emph {et~al.}(2013)\citenamefont {Haazen},
  \citenamefont {Murè}, \citenamefont {Franken}, \citenamefont {Lavrijsen},
  \citenamefont {Swagten},\ and\ \citenamefont {Koopmans}}]{Haazen2013}%
  \BibitemOpen
  \bibfield  {author} {\bibinfo {author} {\bibfnamefont {P.~P.~J.}\
  \bibnamefont {Haazen}}, \bibinfo {author} {\bibfnamefont {E.}~\bibnamefont
  {Murè}}, \bibinfo {author} {\bibfnamefont {J.~H.}\ \bibnamefont {Franken}},
  \bibinfo {author} {\bibfnamefont {R.}~\bibnamefont {Lavrijsen}}, \bibinfo
  {author} {\bibfnamefont {H.~J.~M.}\ \bibnamefont {Swagten}}, \ and\ \bibinfo
  {author} {\bibfnamefont {B.}~\bibnamefont {Koopmans}},\ }\href
  {http://dx.doi.org/10.1038/nmat3553} {\bibfield  {journal} {\bibinfo
  {journal} {Nat Mater}\ }\textbf {\bibinfo {volume} {12}},\ \bibinfo {pages}
  {299} (\bibinfo {year} {2013})}\BibitemShut {NoStop}%
\bibitem [{\citenamefont {Emori}\ \emph {et~al.}(2013)\citenamefont {Emori},
  \citenamefont {Bauer}, \citenamefont {Ahn}, \citenamefont {Martinez},\ and\
  \citenamefont {Beach}}]{Emori2013}%
  \BibitemOpen
  \bibfield  {author} {\bibinfo {author} {\bibfnamefont {S.}~\bibnamefont
  {Emori}}, \bibinfo {author} {\bibfnamefont {U.}~\bibnamefont {Bauer}},
  \bibinfo {author} {\bibfnamefont {S.-M.}\ \bibnamefont {Ahn}}, \bibinfo
  {author} {\bibfnamefont {E.}~\bibnamefont {Martinez}}, \ and\ \bibinfo
  {author} {\bibfnamefont {G.~S.~D.}\ \bibnamefont {Beach}},\ }\href
  {http://dx.doi.org/10.1038/nmat3675} {\bibfield  {journal} {\bibinfo
  {journal} {Nat Mater}\ }\textbf {\bibinfo {volume} {12}},\ \bibinfo {pages}
  {611} (\bibinfo {year} {2013})}\BibitemShut {NoStop}%
\bibitem [{\citenamefont {Torrejon}\ \emph {et~al.}(2014)\citenamefont
  {Torrejon}, \citenamefont {Kim}, \citenamefont {Sinha}, \citenamefont
  {Mitani}, \citenamefont {Hayashi}, \citenamefont {Yamanouchi},\ and\
  \citenamefont {Ohno}}]{Torrejon2014}%
  \BibitemOpen
  \bibfield  {author} {\bibinfo {author} {\bibfnamefont {J.}~\bibnamefont
  {Torrejon}}, \bibinfo {author} {\bibfnamefont {J.}~\bibnamefont {Kim}},
  \bibinfo {author} {\bibfnamefont {J.}~\bibnamefont {Sinha}}, \bibinfo
  {author} {\bibfnamefont {S.}~\bibnamefont {Mitani}}, \bibinfo {author}
  {\bibfnamefont {M.}~\bibnamefont {Hayashi}}, \bibinfo {author} {\bibfnamefont
  {M.}~\bibnamefont {Yamanouchi}}, \ and\ \bibinfo {author} {\bibfnamefont
  {H.}~\bibnamefont {Ohno}},\ }\href {http://dx.doi.org/10.1038/ncomms5655}
  {\bibfield  {journal} {\bibinfo  {journal} {Nat Commun}\ }\textbf {\bibinfo
  {volume} {5}},\  (\bibinfo {year} {2014})}\BibitemShut {NoStop}%
\bibitem [{\citenamefont {Chen}\ \emph {et~al.}(2013)\citenamefont {Chen},
  \citenamefont {Takahashi}, \citenamefont {Nakayama}, \citenamefont
  {Althammer}, \citenamefont {Goennenwein}, \citenamefont {Saitoh},\ and\
  \citenamefont {Bauer}}]{Chen2013}%
  \BibitemOpen
  \bibfield  {author} {\bibinfo {author} {\bibfnamefont {Y.-T.}\ \bibnamefont
  {Chen}}, \bibinfo {author} {\bibfnamefont {S.}~\bibnamefont {Takahashi}},
  \bibinfo {author} {\bibfnamefont {H.}~\bibnamefont {Nakayama}}, \bibinfo
  {author} {\bibfnamefont {M.}~\bibnamefont {Althammer}}, \bibinfo {author}
  {\bibfnamefont {S.~T.~B.}\ \bibnamefont {Goennenwein}}, \bibinfo {author}
  {\bibfnamefont {E.}~\bibnamefont {Saitoh}}, \ and\ \bibinfo {author}
  {\bibfnamefont {G.~E.~W.}\ \bibnamefont {Bauer}},\ }\href {\doibase
  10.1103/PhysRevB.87.144411} {\bibfield  {journal} {\bibinfo  {journal} {Phys.
  Rev. B}\ }\textbf {\bibinfo {volume} {87}},\ \bibinfo {pages} {144411}
  (\bibinfo {year} {2013})}\BibitemShut {NoStop}%
\bibitem [{\citenamefont {Nakayama}\ \emph {et~al.}(2013)\citenamefont
  {Nakayama}, \citenamefont {Althammer}, \citenamefont {Chen}, \citenamefont
  {Uchida}, \citenamefont {Kajiwara}, \citenamefont {Kikuchi}, \citenamefont
  {Ohtani}, \citenamefont {Gepr\"ags}, \citenamefont {Opel}, \citenamefont
  {Takahashi}, \citenamefont {Gross}, \citenamefont {Bauer}, \citenamefont
  {Goennenwein},\ and\ \citenamefont {Saitoh}}]{Nakayama.PRL.2013}%
  \BibitemOpen
  \bibfield  {author} {\bibinfo {author} {\bibfnamefont {H.}~\bibnamefont
  {Nakayama}}, \bibinfo {author} {\bibfnamefont {M.}~\bibnamefont {Althammer}},
  \bibinfo {author} {\bibfnamefont {Y.-T.}\ \bibnamefont {Chen}}, \bibinfo
  {author} {\bibfnamefont {K.}~\bibnamefont {Uchida}}, \bibinfo {author}
  {\bibfnamefont {Y.}~\bibnamefont {Kajiwara}}, \bibinfo {author}
  {\bibfnamefont {D.}~\bibnamefont {Kikuchi}}, \bibinfo {author} {\bibfnamefont
  {T.}~\bibnamefont {Ohtani}}, \bibinfo {author} {\bibfnamefont
  {S.}~\bibnamefont {Gepr\"ags}}, \bibinfo {author} {\bibfnamefont
  {M.}~\bibnamefont {Opel}}, \bibinfo {author} {\bibfnamefont {S.}~\bibnamefont
  {Takahashi}}, \bibinfo {author} {\bibfnamefont {R.}~\bibnamefont {Gross}},
  \bibinfo {author} {\bibfnamefont {G.~E.~W.}\ \bibnamefont {Bauer}}, \bibinfo
  {author} {\bibfnamefont {S.~T.~B.}\ \bibnamefont {Goennenwein}}, \ and\
  \bibinfo {author} {\bibfnamefont {E.}~\bibnamefont {Saitoh}},\ }\href
  {\doibase 10.1103/PhysRevLett.110.206601} {\bibfield  {journal} {\bibinfo
  {journal} {Phys. Rev. Lett.}\ }\textbf {\bibinfo {volume} {110}},\ \bibinfo
  {pages} {206601} (\bibinfo {year} {2013})}\BibitemShut {NoStop}%
\bibitem [{\citenamefont {Althammer}\ \emph {et~al.}(2013)\citenamefont
  {Althammer}, \citenamefont {Meyer}, \citenamefont {Nakayama}, \citenamefont
  {Schreier}, \citenamefont {Altmannshofer}, \citenamefont {Weiler},
  \citenamefont {Huebl}, \citenamefont {Gepr\"ags}, \citenamefont {Opel},
  \citenamefont {Gross}, \citenamefont {Meier}, \citenamefont {Klewe},
  \citenamefont {Kuschel}, \citenamefont {Schmalhorst}, \citenamefont {Reiss},
  \citenamefont {Shen}, \citenamefont {Gupta}, \citenamefont {Chen},
  \citenamefont {Bauer}, \citenamefont {Saitoh},\ and\ \citenamefont
  {Goennenwein}}]{Althammer.PRB.2013}%
  \BibitemOpen
  \bibfield  {author} {\bibinfo {author} {\bibfnamefont {M.}~\bibnamefont
  {Althammer}}, \bibinfo {author} {\bibfnamefont {S.}~\bibnamefont {Meyer}},
  \bibinfo {author} {\bibfnamefont {H.}~\bibnamefont {Nakayama}}, \bibinfo
  {author} {\bibfnamefont {M.}~\bibnamefont {Schreier}}, \bibinfo {author}
  {\bibfnamefont {S.}~\bibnamefont {Altmannshofer}}, \bibinfo {author}
  {\bibfnamefont {M.}~\bibnamefont {Weiler}}, \bibinfo {author} {\bibfnamefont
  {H.}~\bibnamefont {Huebl}}, \bibinfo {author} {\bibfnamefont
  {S.}~\bibnamefont {Gepr\"ags}}, \bibinfo {author} {\bibfnamefont
  {M.}~\bibnamefont {Opel}}, \bibinfo {author} {\bibfnamefont {R.}~\bibnamefont
  {Gross}}, \bibinfo {author} {\bibfnamefont {D.}~\bibnamefont {Meier}},
  \bibinfo {author} {\bibfnamefont {C.}~\bibnamefont {Klewe}}, \bibinfo
  {author} {\bibfnamefont {T.}~\bibnamefont {Kuschel}}, \bibinfo {author}
  {\bibfnamefont {J.-M.}\ \bibnamefont {Schmalhorst}}, \bibinfo {author}
  {\bibfnamefont {G.}~\bibnamefont {Reiss}}, \bibinfo {author} {\bibfnamefont
  {L.}~\bibnamefont {Shen}}, \bibinfo {author} {\bibfnamefont {A.}~\bibnamefont
  {Gupta}}, \bibinfo {author} {\bibfnamefont {Y.-T.}\ \bibnamefont {Chen}},
  \bibinfo {author} {\bibfnamefont {G.~E.~W.}\ \bibnamefont {Bauer}}, \bibinfo
  {author} {\bibfnamefont {E.}~\bibnamefont {Saitoh}}, \ and\ \bibinfo {author}
  {\bibfnamefont {S.~T.~B.}\ \bibnamefont {Goennenwein}},\ }\href {\doibase
  10.1103/PhysRevB.87.224401} {\bibfield  {journal} {\bibinfo  {journal} {Phys.
  Rev. B}\ }\textbf {\bibinfo {volume} {87}},\ \bibinfo {pages} {224401}
  (\bibinfo {year} {2013})}\BibitemShut {NoStop}%
\bibitem [{\citenamefont {Avci}\ \emph {et~al.}(2015)\citenamefont {Avci},
  \citenamefont {Garello}, \citenamefont {Ghosh}, \citenamefont {Gabureac},
  \citenamefont {Alvarado},\ and\ \citenamefont {Gambardella}}]{Avci2015}%
  \BibitemOpen
  \bibfield  {author} {\bibinfo {author} {\bibfnamefont {C.~O.}\ \bibnamefont
  {Avci}}, \bibinfo {author} {\bibfnamefont {K.}~\bibnamefont {Garello}},
  \bibinfo {author} {\bibfnamefont {A.}~\bibnamefont {Ghosh}}, \bibinfo
  {author} {\bibfnamefont {M.}~\bibnamefont {Gabureac}}, \bibinfo {author}
  {\bibfnamefont {S.~F.}\ \bibnamefont {Alvarado}}, \ and\ \bibinfo {author}
  {\bibfnamefont {P.}~\bibnamefont {Gambardella}},\ }\href
  {http://dx.doi.org/10.1038/nphys3356} {\bibfield  {journal} {\bibinfo
  {journal} {Nat Phys}\ }\textbf {\bibinfo {volume} {11}},\ \bibinfo {pages}
  {570} (\bibinfo {year} {2015})}\BibitemShut {NoStop}%
\bibitem [{\citenamefont {Kim}\ \emph {et~al.}()\citenamefont {Kim},
  \citenamefont {Sheng}, \citenamefont {Takahashi}, \citenamefont {Mitani},\
  and\ \citenamefont {Hayashi}}]{Kim.Arxiv.2015}%
  \BibitemOpen
  \bibfield  {author} {\bibinfo {author} {\bibfnamefont {J.}~\bibnamefont
  {Kim}}, \bibinfo {author} {\bibfnamefont {P.}~\bibnamefont {Sheng}}, \bibinfo
  {author} {\bibfnamefont {S.}~\bibnamefont {Takahashi}}, \bibinfo {author}
  {\bibfnamefont {S.}~\bibnamefont {Mitani}}, \ and\ \bibinfo {author}
  {\bibfnamefont {M.}~\bibnamefont {Hayashi}},\ }\href@noop {} {\enquote
  {\bibinfo {title} {Giant spin hall magnetoresistance in metallic bilayers},}\
  }\Eprint {http://arxiv.org/abs/arXiv:1503.08903} {arXiv:1503.08903}
  \BibitemShut {NoStop}%
\bibitem [{\citenamefont {Kim}\ \emph {et~al.}(2013)\citenamefont {Kim},
  \citenamefont {Sinha}, \citenamefont {Hayashi}, \citenamefont {Yamanouchi},
  \citenamefont {Fukami}, \citenamefont {Suzuki}, \citenamefont {Mitani},\ and\
  \citenamefont {Ohno}}]{Kim2013}%
  \BibitemOpen
  \bibfield  {author} {\bibinfo {author} {\bibfnamefont {J.}~\bibnamefont
  {Kim}}, \bibinfo {author} {\bibfnamefont {J.}~\bibnamefont {Sinha}}, \bibinfo
  {author} {\bibfnamefont {M.}~\bibnamefont {Hayashi}}, \bibinfo {author}
  {\bibfnamefont {M.}~\bibnamefont {Yamanouchi}}, \bibinfo {author}
  {\bibfnamefont {S.}~\bibnamefont {Fukami}}, \bibinfo {author} {\bibfnamefont
  {T.}~\bibnamefont {Suzuki}}, \bibinfo {author} {\bibfnamefont
  {S.}~\bibnamefont {Mitani}}, \ and\ \bibinfo {author} {\bibfnamefont
  {H.}~\bibnamefont {Ohno}},\ }\href {http://dx.doi.org/10.1038/nmat3522}
  {\bibfield  {journal} {\bibinfo  {journal} {Nat Mater}\ }\textbf {\bibinfo
  {volume} {12}},\ \bibinfo {pages} {240} (\bibinfo {year} {2013})}\BibitemShut
  {NoStop}%
\bibitem [{\citenamefont {Garello}\ \emph {et~al.}(2013)\citenamefont
  {Garello}, \citenamefont {Miron}, \citenamefont {Avci}, \citenamefont
  {Freimuth}, \citenamefont {Mokrousov}, \citenamefont {Blugel}, \citenamefont
  {Auffret}, \citenamefont {Boulle}, \citenamefont {Gaudin},\ and\
  \citenamefont {Gambardella}}]{Garello2013}%
  \BibitemOpen
  \bibfield  {author} {\bibinfo {author} {\bibfnamefont {K.}~\bibnamefont
  {Garello}}, \bibinfo {author} {\bibfnamefont {I.~M.}\ \bibnamefont {Miron}},
  \bibinfo {author} {\bibfnamefont {C.~O.}\ \bibnamefont {Avci}}, \bibinfo
  {author} {\bibfnamefont {F.}~\bibnamefont {Freimuth}}, \bibinfo {author}
  {\bibfnamefont {Y.}~\bibnamefont {Mokrousov}}, \bibinfo {author}
  {\bibfnamefont {S.}~\bibnamefont {Blugel}}, \bibinfo {author} {\bibfnamefont
  {S.}~\bibnamefont {Auffret}}, \bibinfo {author} {\bibfnamefont
  {O.}~\bibnamefont {Boulle}}, \bibinfo {author} {\bibfnamefont
  {G.}~\bibnamefont {Gaudin}}, \ and\ \bibinfo {author} {\bibfnamefont
  {P.}~\bibnamefont {Gambardella}},\ }\href
  {http://dx.doi.org/10.1038/nnano.2013.145} {\bibfield  {journal} {\bibinfo
  {journal} {Nat Nano}\ }\textbf {\bibinfo {volume} {8}},\ \bibinfo {pages}
  {587} (\bibinfo {year} {2013})}\BibitemShut {NoStop}%
\bibitem [{\citenamefont {N\'{e}el}()}]{Neel1962}%
  \BibitemOpen
  \bibfield  {author} {\bibinfo {author} {\bibfnamefont {L.}~\bibnamefont
  {N\'{e}el}},\ }\href@noop {} {\bibfield  {journal} {\bibinfo  {journal} {Cr.
  Hebd. Acad. Sci.}\ }\textbf {\bibinfo {volume} {255}},\ \bibinfo {pages}
  {1676}}\BibitemShut {NoStop}%
\bibitem [{\citenamefont {Zhang}\ \emph {et~al.}(2015)\citenamefont {Zhang},
  \citenamefont {Fukami}, \citenamefont {Sato}, \citenamefont {Matsukura},\
  and\ \citenamefont {Ohno}}]{Zhang2015}%
  \BibitemOpen
  \bibfield  {author} {\bibinfo {author} {\bibfnamefont {C.}~\bibnamefont
  {Zhang}}, \bibinfo {author} {\bibfnamefont {S.}~\bibnamefont {Fukami}},
  \bibinfo {author} {\bibfnamefont {H.}~\bibnamefont {Sato}}, \bibinfo {author}
  {\bibfnamefont {F.}~\bibnamefont {Matsukura}}, \ and\ \bibinfo {author}
  {\bibfnamefont {H.}~\bibnamefont {Ohno}},\ }\href {\doibase
  http://dx.doi.org/10.1063/1.4926371} {\bibfield  {journal} {\bibinfo
  {journal} {Applied Physics Letters}\ }\textbf {\bibinfo {volume} {107}},\
  \bibinfo {eid} {012401} (\bibinfo {year} {2015})}\BibitemShut {NoStop}%
\end{thebibliography}%

\section*{Acknowledgments}
\label{sec:acknowledgments}

This work was supported by Science Foundation Ireland through AMBER, and from
grant 13/ERC/I2561. KR acknowledges financial support from the European
Community's Seventh Framework Programme IFOX, NMP3-LA-2010-246102.  DB
acknowledges financial support from IRCSET.

The authors thank G.Q. Yu for stimulating this work. We acknowledge G.Q. Yu, N.
Thiyagarajah, and J.Y. Chen for fruitful discussions and Q. Chevigny for assistance
in lithography.

\section*{Author contributions statement}
\label{sec:contributions}
Y.C.L. and D.B. contributed equally to this work. Y.C.L. and D.B. designed the
experiment and
planned the study with the input from K.R. D.B. grew the samples and fabricated
the
devices. Y.C.L. and D.B. measured the devices. D.B. performed data analysis.
Y.C.L. and D.B.
wrote the manuscript with advice from J.M.D.C. and P.S.

\section*{Additional information}

The authors declare no competing financial interests.

\end{document}